\documentclass[aps,prl,floatfix,superscriptaddress]{revtex4}
\usepackage{amsmath,amssymb}
\usepackage{graphicx,float}
\usepackage{times,txfonts}
\usepackage[usenames, dvipsnames]{color}
\usepackage{multirow}
\usepackage{bbold}
\usepackage{color}
\usepackage{soul}
\usepackage{hyperref}
\usepackage{times,txfonts}
\usepackage{natbib}
\usepackage{blindtext}
\usepackage{wasysym}

\begin{document}

\title{Control of spatially rotating structures in diffractive Kerr cavities}

\author{Alison M. Yao, Christopher J. Gibson and Gian-Luca Oppo }

\address{SUPA and Department of Physics, University of Strathclyde, Glasgow G4 0NG, Scotland, E.U.}

\email{alison.yao@strath.ac.uk, g.l.oppo@strath.ac.uk}

\begin{abstract}
Turing patterns in self-focussing nonlinear optical cavities pumped by beams carrying orbital angular momentum (OAM) $m$ are shown to rotate with an angular velocity $\omega = 2m/R^2$ on rings of radii $R$. We verify this prediction in 1D models on a ring and for 2D Laguerre-Gaussian and top-hat pumps with OAM. Full control over the angular velocity of the pattern in the range $- 2m/R^2 \le \omega \le 2m/R^2$ is obtained by using cylindrical vector beam pumps that consist of orthogonally polarized eigenmodes with equal and opposite OAM. Using Poincar\'e beams that consist of orthogonally polarized eigenmodes with different magnitudes of OAM, the resultant angular velocity is  $\omega = (m_L + m_R)/R^2$, where $m_L, m_R$ are the OAMs of the eigenmodes, assuming good overlap between the eigenmodes. If there is no, or very little, overlap between the modes then concentric Turing pattern rings, each with angular velocity $\omega = 2m_{L,R}/R^2$ will result. This can lead to, for example, concentric, counter-rotating Turing patterns creating an 'optical peppermill'-type structure. Full control over the speeds of multiple rings has potential applications in particle manipulation and stretching, atom trapping, and circular transport of cold atoms and BEC wavepackets.
\end{abstract}

\maketitle

\section{Introduction}
Pattern formation is ubiquitous in nonlinear dynamical systems, the most famous example being Turing patterns in reaction-diffusion systems \cite{Turing52}. In optical cavities, spontaneous spatial pattern formation results due to the interplay of a nonlinearity and a spatial coupling, such as diffraction or dispersion. Such systems are very well described by the Lugiato-Lefever equation \cite{LL87}. 
In this paper we consider the effect of pumping optical cavities containing a self-focussing Kerr medium with beams carrying optical angular momentum (OAM) and show the formation of rotating Turing structures. We derive analytical expressions that fully describe these two-dimensional rotating Turing structures in single field (scalar) Kerr resonators and confirm our predictions numerically using pumps consisting of Laguerre-Gaussian modes or `'top-hats'' carrying OAM. In particular, we show that the angular velocity $\omega$ of the patterns is fully determined by the OAM $m$ of the pump and the radius $R$ of the ring structure according to $\omega = 2 m/R^2$. Spatial structures rotating on a transverse ring including cavity solitons can be considered as slow light pulses with fully controllable speed and structure for use in optical quantum memories and delay lines. These studies complete early investigations that focused on optical parametric oscillators, semiconductor heterostructures and photorefractive materials, respectively \cite{Oppo01,Kheradmand03,Caullet12}.

Fully-structured light consisting of a vector superposition of two scalar OAM-carrying Laguerre-Gaussian eigenmodes with orthogonal circular polarizations \cite{Zhan09,Beckley10,Galvez12}, has attracted increasing attention for a number of applications \cite{Nesterov00,Novotny01,Dorn03,Bouchard16}. The inclusion of a second field component in the light-matter interaction inside the cavity offers further degrees of control in the shape and polarization of the pump and the resultant nonlinear structures. In particular we show how the use of fully-structured light to pump the cavity allows us full control over the angular velocity of the Turing structures. Using numerical simulations we demonstrate how biasing cylindrical vector (CV) beam pumps - orthogonally polarized eigenmodes with equal and opposite OAM ($m_L = -m_R$) - allows us to produce patterns with angular velocity $-2 m/R^2 \le \omega \le 2 m/R^2$ and that Poincar{\'e} pumps - orthogonally polarized eigenmodes with different magnitudes of OAM - produce patterns with angular velocity $\omega = (m_L + m_R)/R^2$. Applications of these rotating structures to particle manipulation, optical beam shaping and photonic devices will is discussed.  Finally we give examples of fields with counter-rotating Turing patterns, the `'optical peppermill'', that may be of particular interest in trapping, manipulating and deforming biological specimens.

\section{The Lugiato-Lefever Case}
We start with the description of Kerr media in optical cavities through the well-known Lugiato-Lefever equation (LLE) in two transverse dimensions \cite{LL87}:
\begin{equation}
\label{eqn:LL2D}
\partial_t E = P-(1+i\theta) E + i \beta |E|^2 E + i \nabla^2 E
\end{equation}
where $E$ is the intracavity field, $P$ is the amplitude of the input pump, $\theta$ is the detuning between the input pump and the closest cavity resonance, $\beta$ is proportional to the Kerr coefficient of the nonlinear material, and the term with the transverse Laplacian $\nabla^2$ describes diffraction and can be written in either Cartesian or polar coordinates. 
The time scale has been normalised by $\tau_p$ the mean lifetime of photons in the cavity given by $2L/cT$ for a unidirectional ring cavity and by $4L/cT$ for a Fabry-Perot cavity, with $L$ being the cavity length, $T$ the (intensity) transmission coefficient of the cavity mirrors and $c$ the speed of light in vacuum. The transverse spatial scale ($x$, $y$) has been normalised by 
\begin{equation}
\label{spacenorm}
a = \left( \frac{c \tau_p}{2 K} \right)^{1/2} = \left( \frac{c \lambda \tau_p}{4 \pi } \right)^{1/2}  
= \left( \frac{L}{k T} \right)^{1/2} 
\end{equation}
where $k$ and $\lambda$ are the wavevector and wavelength of the input light, respectively. \\

In order to derive some analytical results, we note that LG modes with $m > 0$ can be considered as rings of fixed radius $R_m$ normalised via (\ref{spacenorm}). We therefore express the transverse Laplacian in polar coordinates $(R, \varphi)$ and, as $R$ can be considered a constant, we can write the LLE Eq. (\ref{eqn:LL2D}) in one angular transverse dimension:
\begin{equation}
\label{LL1D}
\partial_t E = P-(1+i\theta) E + i \beta |E|^2 E + \frac{i}{R^2} \frac{\partial^2 E}{\partial \varphi^2}  \; .
\end{equation}

As the focus of this work is the effect of pumping the ring with light carrying orbital angular momentum (OAM), we consider pumps of the form:
\begin{equation}
\label{pump}
P = P_m e^{i m \varphi}
\end{equation}
where $P_m$ is a complex amplitude independent of $\varphi$, and $m$ is an integer corresponding to the topological charge of the optical vortex. In this case we consider solutions of the form:
\begin{equation}
\label{Fsolut}
E(\varphi,t)=F(\varphi,t)  e^{i m \varphi} \; 
\end{equation}
that satisfy the equation:
\begin{eqnarray}
\label{LL1DF}
\frac{\partial F}{\partial t} = P_m-\left[ 1 + i \left( \theta + \frac{m^2}{R^2} \right) \right] F 
+ i \beta |F|^2 F
- \frac{2m}{R^2}  \frac {\partial F}{\partial \varphi} + \frac{i} {R^2} \frac{\partial^2 F}{\partial \varphi^2} \; .
\end{eqnarray}
One effect of the OAM-dependent solution (\ref{Fsolut}) is that the detuning is modified by an amount $m^2/R^2$. We note that this phase shift is independent of the sign of OAM (i.e. left- or right-hand phase circulation) and the overall effect is to increase the cavity off-tuning for positive $\theta$ and to (partially) compensate the detuning in the case of negative $\theta$. Moreover, this OAM-dependent detuning increases when the radius of the ring decreases. 

\subsection{Homogeneous Stationary States}
For any value of $m$, the homogenenous stationary solutions $F_s$ are obtained from:
\begin{equation}
\label{SSpump}
P_m = F_s \left[ 1 + i \left( \theta + \frac{m^2}{R^2} - \beta I_s \right) \right]  
\end{equation}
where  $I_s$ is the intensity of the stationary solution $I_s=|F_s|^2=|E_s|^2$.
%
%
Once the stationary intensity $I_s$ is selected, the amplitude and phase of the pump field are obtained from (\ref{SSpump}) implicitly. 
We highlight the radial dependent detuning term that comes from the OAM associated with the helical phase of the stationary solution (\ref{Fsolut}) and note that for $m\neq0$, homogeneous stationary states of Eq. (\ref{LL1DF}) correspond to stationary states for the field $E$ that are not homogeneous in the phase $\varphi$.

\subsection{Turing instabilities on the ring: $m=0$ }
For $m=0$, $F=E$ and Eq. (\ref{LL1DF}) is equivalent to Eq. (\ref{LL1D}). Both Eqs. (\ref{eqn:LL2D}) and (\ref{LL1D}) are well known to display a Turing instability of the homogeneous stationary state. In order to analyze the stability of the solutions we introduce a small perturbation with wavevector $k$: $E = E_s e^{\lambda(k)}\delta E$ and neglect terms nonlinear in $\delta E, \delta E^*$. Performing a linear stability analysis (LSA) we find that 
above the Turing instability, both Eqs. (\ref{eqn:LL2D}) and (\ref{LL1D}) with $m=0$ experience the linear growth of a perturbation with wavevector $k$ given by \cite{LL87,evalue} :
\begin{equation}
\label{turing}
\lambda(k) = -1 \pm \sqrt{4 \Delta \beta I_s - 3 \beta^2 I_s^2 - \Delta^2  } \; .
\end{equation}
%
where $\Delta = \theta + k^2$. It is clear from Eq. (\ref{turing}) that if the square root is imaginary, there are no instabilities since both eigenvalues have negative real part. For a real eigenvalue to be positive, the quantity in the square root has to be larger than one. The instability boundary where the square root in (\ref{turing}) is exactly equal to one, provides a relation between the detuning $\Delta$ (which contains the wavevector $k$) and the stationary intensity $I_s$:
\begin{equation}
\label{turing2}
\Delta = 2 \beta I_s \pm \sqrt{\beta^2 I_s^2 - 1} \; .
\end{equation}
This shows that there is an instability threshold in the stationary intensity given by $I_s^c=1/\beta$. For a given $I_s>I_s^c=1/\beta$ the most unstable wavevector is obtained by finding the maximum of the square root in (\ref{turing}) when changing $\Delta$:
\begin{equation}
\label{kc}
k_c 
= \sqrt{2 \beta I_s - \theta} \, .
\end{equation}
Above threshold, $N$ peaks appear along the ring separated by a distance given by, or close to, the wavelength of the Turing structure $\Lambda_c = 2 \pi / k_c$. Note that for a ring of circumference $2 \pi R$, the number of peaks is $N = 2 \pi R/\Lambda_c = R \sqrt{2 \beta I_s - \theta}$, 
and exactly $N$ peaks fit inside a ring of radius $R$ to satisfy the periodic boundary conditions. 

\subsection{Rotating solutions: $m\neq0$}
We now consider the case of pumps carrying OAM, i.e. $m \neq 0$. Above the instability threshold these are seen numerically to form patterns that move around the ring at a constant angular velocity.
We start by rearranging Eq. (\ref{LL1DF}) such that the first order derivatives are on the l.h.s.:
\begin{eqnarray}
\label{LL1DFv2}
\frac{\partial F}{\partial t} + \frac{2m}{R^2}  \frac {\partial F}{\partial \varphi} = P_m-\left[ 1 + i \left( \theta + \frac{m^2}{R^2} \right) \right] F 
+ i \beta |F|^2 F +  \frac{i} {R^2} \frac{\partial^2 F}{\partial \varphi^2} \; .
\end{eqnarray}
%
Note that this is the generalization of the analysis of a tilted wave front \cite{firthScroggie96} to polar coordinates on a ring. 
We then consider travelling wave solutions to Eq. (\ref{LL1DFv2}) of the form $F(q)$ that depend on the variables $\varphi$ and $t$ through
\begin{equation}
q = \varphi - \omega \, t \; ,
\end{equation}
where $\omega$ is the angular velocity. 
In this case we can write the l.h.s. of Eq. (\ref{LL1DFv2}) as
\begin{eqnarray}
\label{LL1DFv2lhs}
\frac{\partial F}{\partial t} + \frac{2m}{R^2}  \frac {\partial F}{\partial \varphi} = \frac{\partial F(q)}{\partial q} 
\left( -\omega + \frac{2m}{R^2} \right)   \; .
\end{eqnarray}
Clearly this equals zero when
\begin{equation}
\omega = \frac{2m}{R^2} \, ,
\label{eqn:omega}
\end{equation}
and thus there exist rotating solutions $F(q)$ with angular velocity $\omega = 2m/R^2$ that can be determined via 
\begin{eqnarray}
\label{RotSolut}
P_m = \left[ 1 + i \left( \theta + \frac{m^2}{R^2} \right) \right] F - i \beta |F|^2 F -  \frac{i} {R^2} \frac{\partial^2 F}{\partial q^2} \; .
\end{eqnarray}

Apart from a renormalization of the detuning, Eq. (\ref{RotSolut}) is equivalent to the stationary solutions of Eqs. (\ref{LL1DF}) and (\ref{LL1D}) for $m=0$. This means that all the results of the $m=0$ case can be applied to the $m \neq 0$ case starting from the trivial homogeneous state that we have already seen in Eq. (\ref{SSpump}). These travelling wave solutions are equivalent to stationary solutions found using the retarded time $\tau = t - \varphi/\omega$, as described in Appendix A.

Among these travelling waves solutions we can identify Turing patterns for $m \neq 0$ arising from Eq. (\ref{RotSolut}) by analogy with the $m=0$ case. Travelling wave Turing patterns are rotating solutions of Eq. (\ref{LL1DFv2}) satisfying (\ref{turing}) upon the redefinition of the space-dependent detuning
\begin{equation}
\Delta = \theta + \frac{m^2}{R^2} + k^2 
\end{equation}
and critical wavevector:
\begin{equation}
\label{eqn:kcv2}
k_c = 
\sqrt{2 \beta I_s - \theta - \frac{m^2}{R^2}} \, .
\end{equation}
The wavelength of the Turing structure $\Lambda_c = 2 \pi / k_c$ and the number of peaks 
\begin{equation}
N = 2 \pi R/\Lambda_c = R \sqrt{2 \beta I_s - \theta - \frac{m^2}{R^2}}
\label{eqn:NpeaksOAM}
\end{equation}
now both depend on the OAM and the radius of the ring.
Note that for detuning $\theta$ different from $2 \beta I_s$ and for $m^2/R^2$ small, e.g. for small magnitudes of OAM  and large radii, the critical wavevectors from (\ref{kc}) and (\ref{eqn:kcv2}), and hence the number of peaks, are approximately the same. 
Historically, expressions for the angular velocity similar to (\ref{eqn:omega}) had been obtained and applied to rotating domain walls in optical parametric oscillators \cite{Oppo01} and used as numerical ansatz for self-trapped necklace-ring beams in a self-focusing nonlinear Schr\"odinger equation \cite{soljacic01}. 

The present analysis is confined to Turing patterns close to the threshold of instability of the homogeneous stationary state. In Appendix B we show that it is possible to obtain equations that describe rotating Turing patterns for generic values of the pump $P_m$, detuning and OAM.

\section{Numerical Simulations}
Althought the analysis in the previous section assumed a quasi-1D geometry (rings of fixed radius), all of our numerical simulations are performed in 2D. 
\subsection{Laguerre-Gaussian Pumps}

\begin{figure}[htbp]
\centering\includegraphics[width=7cm]{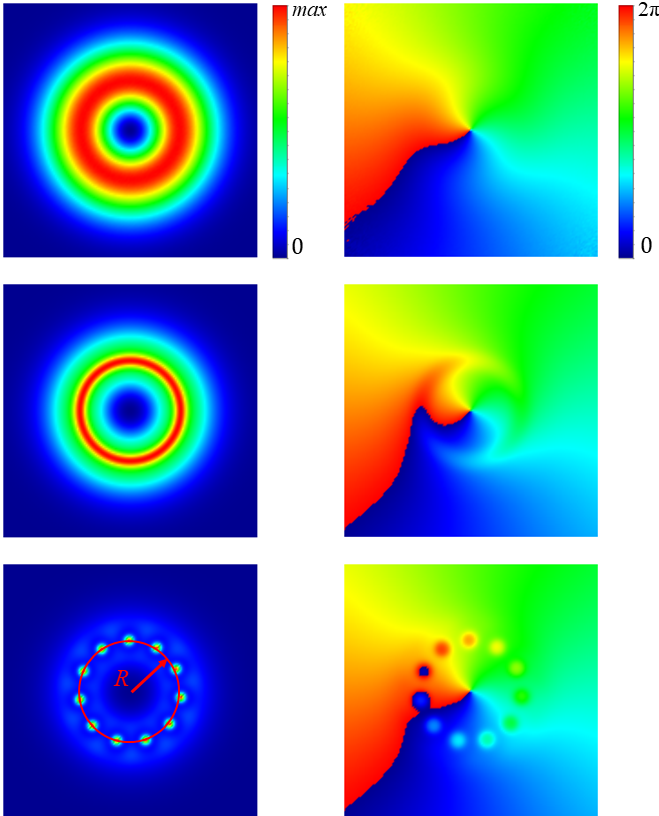}
\caption{
Density plot of intensity (left) and phase (right) during evolution to pattern formation for $m=1$. Parameters are: $I_s = 1.44, \theta = 1, \beta = 2/3, w_0 = 15.0$ (normalised units). 
Top-Bottom: $t = 2, 30, 500$. }
\label{fig:LGoam1-evolv}
\end{figure}

We start by numerically modelling equation \ref{eqn:LL2D} using a Laguerre-Gaussian pump with radial index $p = 0$ \cite{Barnett07}:
\begin{equation}
LG_0^{m} (r, \phi)=\sqrt{\frac{2}{\pi |m| !}} \frac{1}{w_0} \left( \frac{r \sqrt{2}}{w_0} \right)^{|m|} 
 \exp \left( \frac{- r^2}{w_0^2} \right) e^{i m \varphi} = P_m  e^{i m \varphi} ,
\label{eqn:LGmode}
\end{equation}
where $m$ is the OAM and $w_0$ is the beam waist. In Fig.~(\ref{fig:LGoam1-evolv}) we show the time evolution of the field from an LG mode-like ring to a number of bright peaks equally spaced around a ring of maximum intensity.

\begin{figure}[htbp]
\centering\includegraphics[width=9cm]{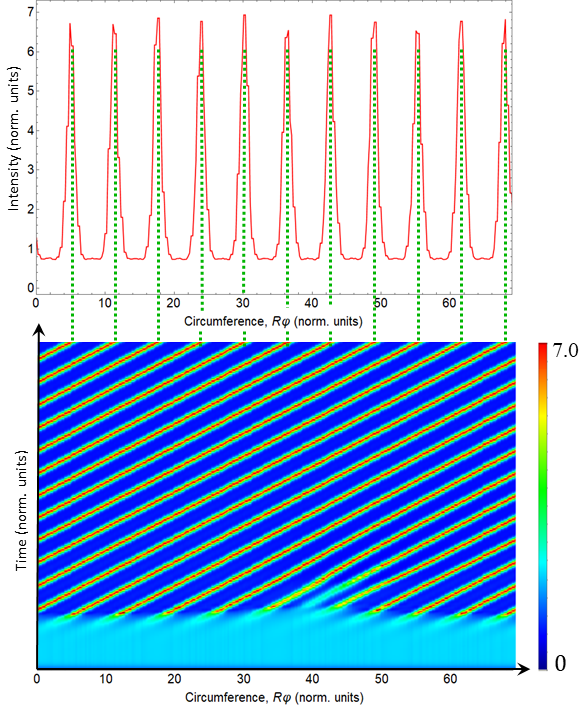}
\caption{
(Top) Intensity of the field at radius $R$ at $t = 1000$ showing 11 peaks.
(Bottom) Time evolution of the intensity at radius $R$ from $t = 0$ to $t = 1000$.
Parameters are: $I_s = 1.44, \theta = 1, \beta = 2/3, w_0 = 15.0, m = 1$.
}
	\label{fig:LGoam1-speed}
\end{figure}

For our given parameters, $I_s = 1.44, \theta = 1, \beta = 2/3, w_0 = 15.0$ (normalised units), we find that $11$ peaks form on a ring of radius $R = 11.0 \pm 0.5$, as shown in the top panel of Fig.~(\ref{fig:LGoam1-speed}). This is in accordance with the closest integer value from our predicted value, using Eq.~(\ref{eqn:NpeaksOAM}), of $10.5 \pm 0.5$. 

From  Eq.~(\ref{eqn:omega}) we predict that the peaks should rotate {\em counter-clockwise} at a constant angular velocity $\omega = 0.0164 \pm 0.0003$. To calculate the angular velocity in the numerical results we plot the time evolution of the field at radius $R$, as shown in the bottom panel of Fig.~(\ref{fig:LGoam1-speed}), where the diagonal red lines correspond to the peaks of intensity. Angular velocity $\omega = \Delta \varphi / \Delta t = \Delta s / (R \Delta t)$ where $\Delta s$ is the distance a peak travels around the circumference of the circle in a time $\Delta t$. Numerically we find $\omega = 0.0164 \pm 0.0003$, in complete agreement with our analytical results. 

\begin{figure}[htbp]
\centering\includegraphics[width=9cm]{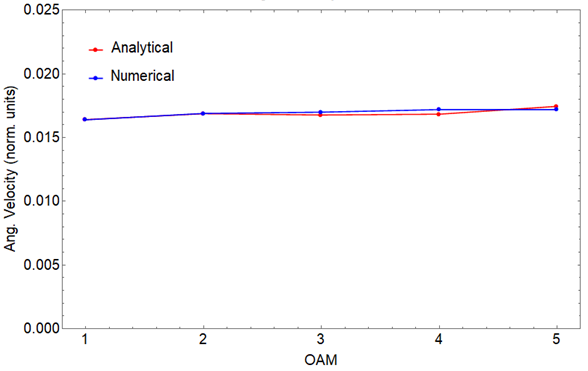}
\caption{
Angular velocity $\omega$ vs OAM index $m$ for LG input pumps.
Parameters are: $I_s = 1.44, \theta = 1, \beta = 2/3, w_0 = 15.0, m = 1 - 5$. 
The blue line corresponds to numerical simulations of Eq. (\ref{eqn:LL2D}), the red line to the analytical result Eq. (\ref{eqn:omega}).
}
\label{fig:lgspeeds-oam}
\end{figure}

We repeated our simulations for an LG pump with $m = -1$, and found that the now peaks rotated {\em clockwise} at the same speed, $\omega = 0.0164 \pm 0.0003$, again as predicted from Eq.~(\ref{eqn:omega}). However, when we measured the angular velocities for $m = 2, 3, 4 \, \& \,5$ numerically we found that in each case $\omega = 0.0164 \pm 0.0003$, as shown by the blue line in Fig.~(\ref{fig:lgspeeds-oam}) i.e. the angular velocity was constant and apparently independent of the OAM of the beam. We can explain this result by noting that the radius of maximum amplitude of the field is OAM-dependent: the red line in Fig.~(\ref{fig:lgspeeds-oam}) shows our numerical results are in good agreement with the predicted angular velocities calculated using the measured values of $R$ for each OAM. 

For any LG mode the radius of maximum amplitude is $r_{max} = w_0 \sqrt{\left| m \right |/2}$.  Substituting this into the angular velocity we find $\omega = 2 m /r^2 = \pm 4/w_0^2$. We therefore expect the angular velocityof LG modes to be independent of $m$ but inversely proportional to the beam waist, $w_0$. 

\begin{figure}[htbp]
\centering\includegraphics[width=9cm]{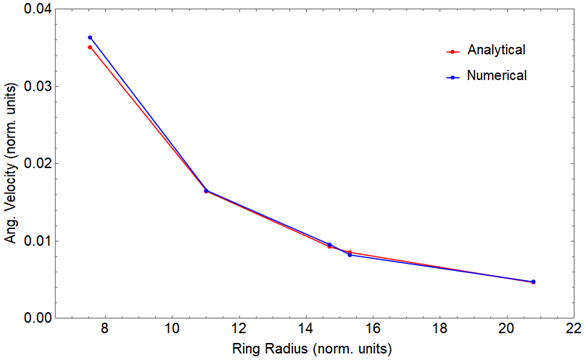}
\caption{
Angular velocity $\omega$ vs the radius of LG input pumps with  $m=1$.
Parameters are: $I_s = 1.44, \theta = 1, \beta = 2/3, w_0 = 10.0-25.0, m = 1$.
The blue line corresponds to numerical simulations of Eq. (\ref{eqn:LL2D}),  
the red line to the analytical result Eq. (\ref{eqn:omega}).
}
\label{fig:lgspeeds-rads}
\end{figure}

To confirm this we numerically integrated Eq.~(\ref{eqn:LL2D}) using LG pumps with $m = 1$ and beam waists of $10.0, 15.0, 20.0$, which formed rings of bright spots at radii $19.23, 28.11, 37.5$, and with a beam waist of $25.0$, which formed two rings of bright spots at radii $39.0, 53.0$. Fig.~(\ref{fig:lgspeeds-rads}) shows that we have very good agreement between the angular velocity of the rings measured numerically (blue line) and the predicted values from Eq.~(\ref{eqn:omega}) using the measured radii (red line). 

%
%

These measurements simultaneously confirm the direct proportionality of the angular velocity to the OAM index $m$ and inverse proportionality to the square of the radius of the input ring (see Eq. (\ref{eqn:omega}). We note that in previous studies of the effect of LG pumps on rotating cavity solitons in semiconductor microresonators \cite{Kheradmand03} and on rotating patterns in photorefractive media in single mirror feedback configurations \cite{Caullet12}, the pump radius and the OAM index where changed simultaneously leading to weak dependencies of the angular velocity on the OAM index (see Fig.~(\ref{fig:lgspeeds-oam}). We believe that both these investigations provide support to the universality of Eq. (\ref{eqn:omega}) when one includes the changing radius of the input pump of LG modes with different OAM.

\subsection{`'Top-hat Pumps carrying OAM''}

We next consider the case where the input pump has a top-hat shape amplitude multiplied by an azimuthal phase:
\begin{equation}
\label{eqn:tophat}
P = \frac{P_m}{2} \left[ 1 - \tanh\left( S(r-r_t) \right) \right] e^{i m \varphi} \,.
\end{equation}
Here $P_m$ is a spatially independent complex amplitude, $S$ and $r_t$ control the steepness of the sides and the radius of the top-hat, respectively, and $m$ is an integer corresponding to the topological charge of the optical vortex, as before. 

\begin{figure}[htbp]
\centering\includegraphics[width=7cm]{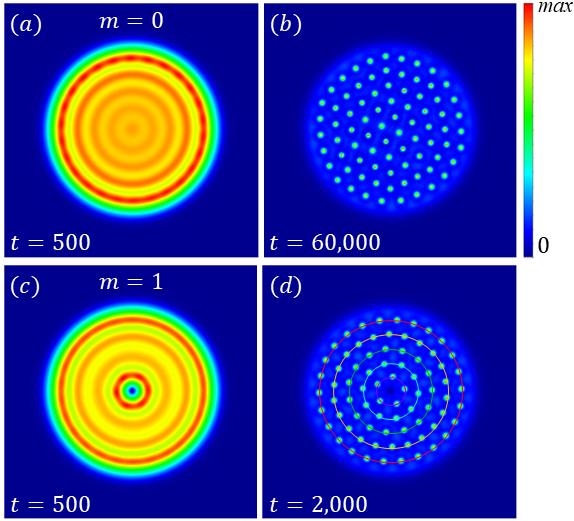}
\caption{
Time evolution of the intensity of a top-hat pump with $m = 0$ (a)-(b) and $m = 1$ (c)-(d). Left-hand-side shows initial development of diffraction rings at $t = 500$, right-hand-side shows the final intensity structure at $t =60,000$ for $m = 0$ and at $t =2,000$ for $m = 1$. When the pump carries OAM (bottom) the diffraction rings have maxima on both outer and inner rings, there is a vortex in the centre, and the Turing patterns are arranged in concentric rings. Parameters are: $I_s = 1.44, \theta = 1, \beta = 2/3, S = 3.0$.
}
\label{fig:th-evolve}
\end{figure}

\begin{figure}[htbp]
\centering\includegraphics[width=7cm]{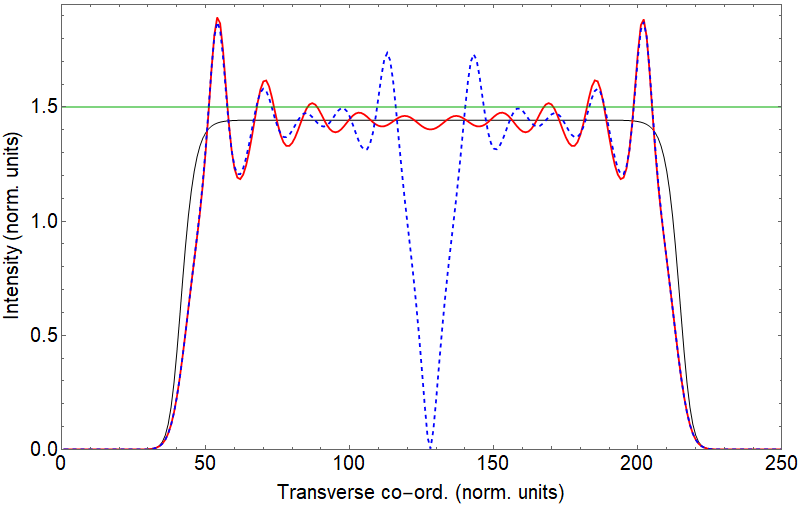}
\caption{
Cross-section of intensity of pump (black) and fields for $m=0$ (red) and $m=1$ (blue) both displaying the radial modulation at the critical wavevector $k_c$. 
Green line is pump intensity threshold for Turing instability. Parameters are: $I_s = 1.44, \theta = 1, \beta = 2/3, S = 3.0, t = 500$.
}
\label{fig:th-steep-kc}
\end{figure}

\begin{figure}[htbp]
\centering\includegraphics[width=10cm]{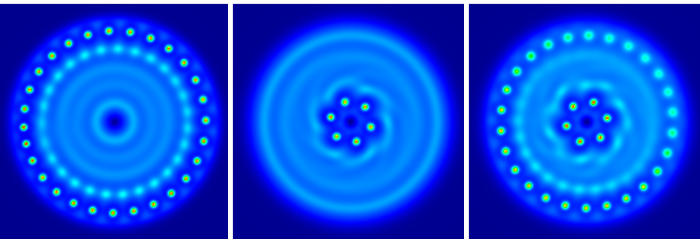}
\caption{
Formation of pattern for top-hat pump with OAM $m = 1$ and steepnesses of $S = 3.0, 1.8, 1.9$ left to right.  For $S = 3.0$ the sides of the pump are steeper than the vortex and so the pattern forms first on the outer ring (left image).  For $S = 1.8$ the sides of the vortex are steeper than the pump and so the pattern forms first on the inner ring (middle image). For $S = 1.9$ the steepnesses are almost balanced and so the pattern can form on the inner and outer rings almost simultaneously (right image).
Parameters are: $I_s = 1.44, \theta = 1, \beta = 2/3$.
}
	\label{fig:th-steep}
\end{figure}

Diffraction due to the finite size of the pump induces concentric rings, whose amplitude decreases from the outer ring inwards for $m=0$, as shown in Fig.~(\ref{fig:th-evolve} (a)). The amplitude of the outermost ring increases with the steepness of the sides of the pump and this can allow its intensity to trigger the Turing instability, see the red line in Fig.~(\ref{fig:th-steep-kc}) in a comparison with $I_s^c=1/\beta$, and an azimuthal pattern forms on the ring. Once the pattern has formed on the outer ring (left panel Fig.~(\ref{fig:th-steep})), we observe a sequence of azimuthal instabilities from the outer to the inner ring. Each patterned ring forms a number of peaks separated by the critical wavelength corresponding to the radius of the particular diffraction ring. The final patterns for $m = 0$ are stationary and close to the centre of the pump they have a hexagonal structure, see Fig.~(\ref{fig:th-evolve} (b)), typical of that found for plane-waves \cite{LL87}. Patterns on the outermost diffraction rings, however, are similar to the well-known daisy or sunflower patterns as observed, for example, in VCSELs with an electronic pump with a steep oxide confinement \cite{Li94, Zhao96, Pereira98, Degen99, Lemke17}. 
When the pump carries OAM, i.e. $m \ne 0$, the phase at the centre of the pump is undefined and hence the field at the origin has to be zero, as is typical for Laguerre-Gaussian modes \cite{yao11}.
The physical effect of this on-axis vortex is the induction of diffractive rings close to the centre of the beam, see the dashed blue line in Fig. (\ref{fig:th-steep-kc}). In this case the intensity of the {\em inner} ring may exceed $I_s^c$ and undergo an azimuthal Turing instability. The distance between the concentric rings is close to the critical wavelength $\Lambda_c$ along the radial direction. Once the Turing pattern has formed on the inner ring we observe a sequence of azimuthal instabilities taking place from the inner to the outer ring. As before, azimuthal peaks are separated by the critical wavelength corresponding to the radial length of the particular diffraction ring, modified by the factor $m^2/R^2$ as shown in Eq.~(\ref{eqn:kcv2}).
Note: The effect of the factor $m^2/R^2$ on the wavelength is stronger towards the centre of the beam where the radius is smaller. Moreover, the size of the central vortex, and hence radius of the first diffraction ring, increases with increasing OAM, $m$. These effects alter the radial wavevector in the vicinity of the vortex and can prevent regular patterns from forming on the innermost rings.

As mentioned above, the steepness of the top-hat pump, determined by $S$ in Eq.~(\ref{eqn:tophat}), and also of the optical vortex, affects the relative intensity of the diffraction rings. 
By careful choice of steepness, and/or the OAM of the pump we can control if the Turing instability first occurs on the inner or the outer ring, or even on both simultaneously.
This is demonstrated in Fig.~(\ref{fig:th-steep}) for a top-hat pump with OAM $m = 1$ and steepnesses of $S = 3.0, 1.8, 1.9$ from left to right. For $S = 3.0$ the sides of the pump are steeper than the vortex and so the pattern forms first on the outer ring (left image).  For $S = 1.8$ the sides of the vortex are steeper than the pump and so the pattern forms first on the inner ring (middle image). For $S = 1.9$ the steepnesses are almost balanced and so the pattern can form on the inner and outer rings almost simultaneously (right image). Ring by ring azimuthal instabilities can occur from the outside to the inside, or vice versa depending on where the pattern is first formed, or may even collide.
\begin{figure}[htbp]
\centering\includegraphics[width=7cm]{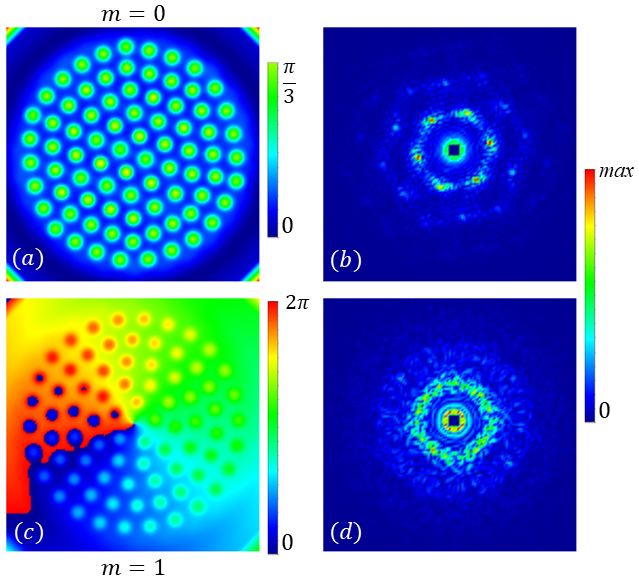}
\caption{
Final phase (a), (c) and final far-field distributions (b), (d) associated with the intensity structures of Fig.~(\ref{fig:th-evolve}) (b) with $m = 0$ and (d) with $m = 1$, respectively. Parameters are: $I_s = 1.44, \theta = 1, \beta = 2/3, S = 3.0$.
}
\label{fig:th-phase}
\end{figure}
%

Whatever the mechanism of ring excitation (innermost or outermost), for $m \ne 0$ Turing patterns appear on the first excited ring and start to rotate before the next ring is excited. This leads to concentric Turing pattern rings where the peaks are seen to rotate at exactly the angular frequency given by (\ref{eqn:omega}) with $R$ equal to the radius of the specific diffraction ring. 
It is interesting to compare the final phase and final far-field distributions for the $m=0$ and $m=1$ cases, as shown in Fig.~(\ref{fig:th-phase}). In both cases there are phase peaks where the local intensity also displays peaks. For $m \neq 0$, the central singularity and the line of phase discontinuity corresponding to the OAM of the input pump are clearly visible. Note that there are no further phase discontinuities in spite of the presence of rotating rings. For $m=0$ the hexagonal structure of the near-field is reflected in the peaks of the far-field distribution (see Fig.~(\ref{fig:th-phase})(b)) while the intensity peaks in the near field for $m \neq 0$ are separated by the critical wavelength $\Lambda_c$ but do not display any specific 2D geometry since they are located on rotating rings (see Fig.~(\ref{fig:th-phase})(d)).
Each ring is decoupled from the rest of the structure, meaning that they behave as independent 1D azimuthal structures although embedded in a fully 2D field. 

\begin{figure}[htbp]
\centering\includegraphics[width=10cm]{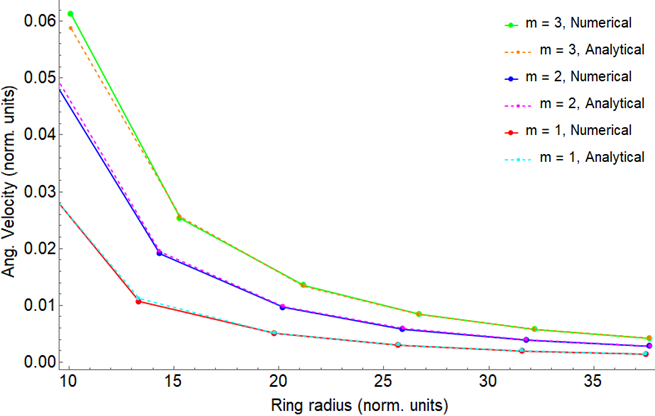}
\caption{
Angular velocity vs ring radius for top-hat pumps with $m = 1$ (red/cyan), $m = 2$ (blue/magenta) and $m = 3$ (green/orange). Solid lines are numerical results, dashed lines are calculated using (\ref{eqn:omega}) with measured radii of rotating rings.
Parameters are: $I_s = 1.44, \theta = 1, \beta = 2/3, S = 3.0$.
}
	\label{fig:th-speeds}
\end{figure}

In Fig.~(\ref{fig:th-speeds}) we plot the angular velocity of each ring that forms versus its radius for top-hat pumps with $m = 1$ (red/cyan), $m = 2$ (blue/magenta) and $m = 3$ (green/orange). The numerical results (solid lines) show excellent agreement with the analytical results (dashed line) calculated using (\ref{eqn:omega}) provided with the measured radii of the rotating rings. This confirms that scalar pumps carrying OAM $m$ form independent Turing patterns on concentric rings of radius $R$ each rotating with constant angular velocity $\omega = 2 m/R^2$. For completion, we note that the dynamics leading to the asymptotic ring rotation for $m \ne 0$ is much faster that that of $m = 0$ where the formation of hexagonal Turing structures is slowed down by the circular symmetry of the input pump. 

%
%

\section{Fully-structured pumps}
Vector, or fully structured light (FSL), beams \cite{Zhan09, Beckley10, Galvez12} have attracted increasing attention for a number of applications. These beams consist of a vector superposition of two scalar orbital angular momentum (OAM) carrying Laguerre-Gaussian (LG) eigenmodes with orthogonal circular polarizations:
\begin{equation}
\vec{E}(r, \phi)=\cos(\gamma) LG_0^{m_L}(r, \phi) \vec{e_l} + e^{i\alpha} \sin(\gamma) LG_0^{m_L}(r, \phi) \vec{e_r},
\label{eqn:FSbeam}    \\
\end{equation}
where $\gamma$ and $\alpha$ give the relative amplitudes and phase, respectively, of the two modes.
We assume throughout that each of the spatial modes takes the form of a Laguerre-Gaussian beam with radial index $p = 0$ as given in (\ref{eqn:LGmode}). The resultant beam has non-uniform spatial intensity, phase and polarization distributions.
To investigate the effect of using an FSL pump we use {\em coupled} Lugiato-Lefever equations \cite{Geddes94}:
\begin{eqnarray}
\label{eqn:LL2Dcoupled}
\partial_t E_{L, R} &=& P_{L, R}-(1+i\theta) E_{L, R} + i \nabla^2 E_{L, R} +  i \beta \left( |E_{L, R}|^2 + 2 |E_{R, L}|^2 \right )E_{L, R} .
\end{eqnarray}
Note that if either $E_L$ or $E_R$ is zero, then the resultant beam is a \emph{scalar} LG mode with spatially uniform right- or left-handed circular polarisation, respectively, and Eq.~(\ref{eqn:LL2Dcoupled}) reduces to the scalar LLE (\ref{eqn:LL2D}) that we have considered so far. 

\subsection{Cylindrical vector beam pumps}
If the two modes have equal but {\it opposite} OAM the resultant beam is know as a \emph{cylindrical vector} (CV) beam~\cite{Zhan09,Galvez12} 
\begin{equation}
\vec{E}(r, \phi)= \cos(\gamma) LG_0^{-m}(r, \phi) \vec{e_l} + e^{i\alpha}  \sin(\gamma) LG_0^{+m}(r, \phi) \vec{e_r}.
\label{eqn:CVbeam}    \\
\end{equation}

\begin{figure}[htbp]
\centering\includegraphics[width=9cm]{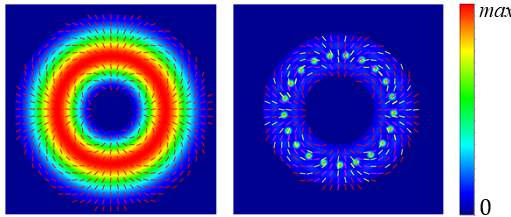}
\caption{
Cylindrical vector beam with OAM $\pm 3$ with transversely structured linear polarization distribution shown by short red lines.
Parameters are: $I_s = 1.44, \theta = 1, \beta = 2/3, w_0 = 15.0, \gamma=\pi/4, \alpha = 0.0$.
}
	\label{fig:CV3}
\end{figure}

If the two modes have equal amplitude ($\gamma = \pi/4$), the pump will have no net OAM and a spatially varying {\em linear} polarization, as shown in Fig.~(\ref{fig:CV3}) for eigenmodes with $|m| = 3$. In this case we find that a pattern of equally-spaced bright spots appears around the ring, as in the scalar case, but this time there is no rotation as the net OAM is zero.
%
%
By changing $\gamma$ in (\ref{eqn:CVbeam}) we change the relative amount of the two eigenmodes, i.e. the bias between the eigenmodes. For $\gamma = 0 \; (\pi/2)$ the pump is a scalar left (right) circularly polarized beam with $-m \, (+m)$ and we find exactly the same behaviour as earlier; in particular, the output Turing pattern rotates at $\omega = \mp 2 m/R^2$. For $\gamma = \pi/4$ the polarization is linear and the Turing pattern is stationary, as mentioned above. For $0 < \gamma < \pi/4$ the left circularly polarized mode dominates.The pump polarization is left ellptical and the output field rotates clockwise (as we would expect for $m < 0$). For $\pi/4 < \gamma < \pi/2$ the right circularly polarized mode dominates.The pump polarization is right ellptical and the output field rotates counter-clockwise (as we would expect for $m > 0$). Note that the polarization structure does not rotate as there is no free propagation \cite{Gibson18}.
\begin{figure}[htbp]
\centering\includegraphics[width=10cm]{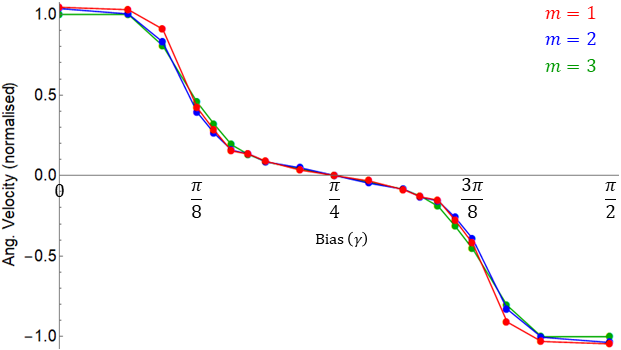}
\caption{
Angular velocity (normalised) vs bias parameter $\gamma$  for cylindrical vector (CV) beams with $|m| = 1$ (red), $|m| = 2$ (blue) and $|m| = 3$ (green). 
Parameters are: $I_s = 1.44, \theta = 1, \beta = 2/3,  w_0 = 15.0, \alpha = 0.0$.
}
	\label{fig:cv-speeds}
\end{figure}

In Fig.~(\ref{fig:cv-speeds}) we plot the angular velocity (normalised to $2 m/R^2$) versus the bias parameter $\gamma$  for cylindrical vector (CV) beams with $|m| = 1$ (red), $|m| = 2$ (blue) and $|m| = 3$ (green). By controlling the bias ($\gamma$) between the two eigenmodes of the CV beam, we can fine tune the angular velocity of the output field from
\begin{equation}
-\frac{2m}{r^2} \le \omega \le \frac{2m}{r^2} .
\end{equation}
%
(Recall that LG modes all have the same angular velocity for any given beam waist $w_0$.)

%
\subsection{Poincar\'e pumps}
\begin{figure}[htbp]
\centering\includegraphics[width=9cm]{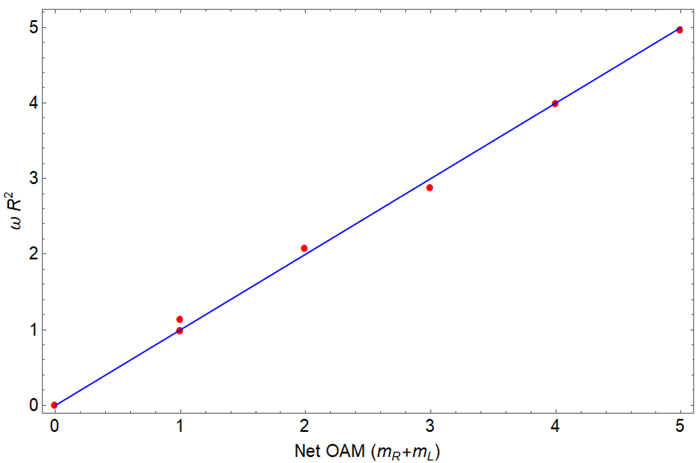}
\caption{
Numerically measured values of $\omega R^2$ against net OAM for different Poincar\'e modes $(m_L, m_R) = (-1,1), (0, 1), (-1, 2), (1, 2), (-1, 3), (1, 3), (-2, 3), (2, 3)$ (red circles). Blue line is $m_L + m_R$.
Parameters are: $I_s = 1.44, \theta = 1, \beta = 2/3, w_0 = 15.0, \gamma=\pi/4, \alpha = 0.0$.
}
	\label{fig:P-speeds}
\end{figure}

 If the two modes have {\em different} magnitudes of OAM, the resultant beam is know as a Poincar\'e beam~\cite{Beckley10}:
\begin{equation}
\vec{E}(r, \phi)= \cos(\gamma) LG_0^{m_L}(r, \phi) \vec{e_l} + e^{i\alpha}  \sin(\gamma) LG_0^{m_R}(r, \phi) \vec{e_r}.
\label{eqn:Pbeam}  
\end{equation}
This carries a net OAM and the polarization can cover all polarization states on the Poincar\'e sphere. In Fig.~(\ref{fig:P-speeds}) we plot the numerically measured values of $\omega R^2$ against net OAM for different Poincar\'e modes $(m_L, m_R) =(-1, 1), (0, 1), (-1, 2), (1, 2), (-1, 3), (1, 3), (-2, 3), (2, 3)$ (red circles). We keep $\gamma = \pi/4, \alpha = 0$. 
We can see that there is very good agreement between our numerical results and the blue line for $m_L + m_R$, suggesting that for Poincar\'e beams, the angular velocity of the output field depends on the net OAM according to:
\begin{equation}
\omega = \frac{m_L + m_R}{R^2} .
\label{eqn:P-speeds}
\end{equation}
%
%
%
%
%
\subsection{Optical `'peppermill''}
\begin{figure}[htbp]
\centering\includegraphics[width=5.5cm]{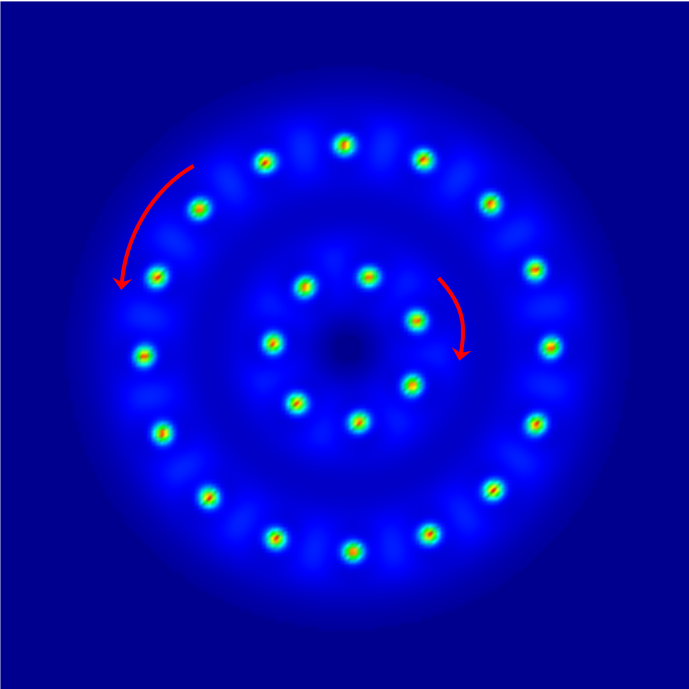}
\caption{
``Optical peppermill'' constructed from orthogonally polarized modes with $m_L = -1$ and $m_R = 8$. The inner ring rotates clockwise while the outer ring rotates counter-clockwise at the same angular velocity.
Parameters are: $I_s = 1.44, \theta = 1, \beta = 2/3, w_0 = 10.0, \gamma=\pi/4, \alpha = 0.0$.  
}
	\label{fig:pepper}
\end{figure}

Up until now we have only considered vector beams with some degree of spatial overlap. By considering eigenmodes with significantly different transverse radii such that there is very little interaction between them we can create, for example, counter-rotating rings of spots such as the ``optical peppermill'' shown in Fig.~(\ref{fig:pepper}). In this case the pump consists of a left -circularly polarized mode with $m_L = -1$ and right-circularly polarized mode with $m_R = 8$. The output field has two rings of peaks, with the outer ring rotating counter-clockwise and the inner clockwise with the same angular velocity  $\omega = 0.0375\pm 0.0015$.
%
%
%
In principle we can produce even more complex superpositions of modes as shown:
\begin{eqnarray}
\vec{E}(r, \phi) & = & \cos(\gamma)E_L(r, \phi) \vec{e_l} + e^{i\alpha}  \sin(\gamma) E_R(r, \phi) \vec{e_r}; \nonumber \\
E_L(r, \phi) & = &  \sum_{i=0}^{n_L} \frac{A_i LG_{m_i}}{\sqrt{\sum_{i=0}^{n_L} A_i^2}}  ; E_R(r, \phi) = \sum_{j=0}^{n_R} \frac{B_j LG_{m_j}}{\sqrt{\sum_{j=0}^{n_R} B_j^2}} ,
\label{eqn:ComPbeam}  
\end{eqnarray}
where $A_i$ and $B_j$ are the contributions of left- and right-circularly polarized LG modes with OAM $m_i, m_j$, respectively. 
This can, for example, allow us to produce the peppermill-type beam shown in Fig.~(\ref{fig:pepper}) but with full and individual control over the speeds of each of the rings simply by biasing each with modes of opposite OAM and orthogonal polarisation, as in Fig.~(\ref{fig:cv-speeds}).
%
%
\section{Conclusion}
We have demonstrated formation and rotation of spatio-temporal patterns in self-focussing nonlinear optical cavities pumped by beams carrying orbital angular momentum, $m$. For scalar pumps we see the formation of a ring, or concentric rings, around an optical vortex that rotate at angular velocity $\omega$. Using a 1D Lugiato-Lefever model we find that $\omega = \frac{2 m}{R^2}$, where $R$ is the radius of each ring. For a 1D azimuthal model this formula is exact but we confirm numerically that these angular velocities extend to the 2D case and demonstrate this using input pumps that are Laguerre-Gaussian modes and `'top-hat'' shaped pumps with OAM.
We note that the radius of maximum intensity of an LG mode scales with the OAM s.t. LG beams with the same beam waist $w_0$ have the same angular velocity, $\omega = \pm 4/w_0^2$. 
This means that we can control the angular velocity of the patterns by the choice of:
\begin{itemize}
\item{OAM, $m$}
\item{beam waist of LG pump, $w_0$}
\item{radius of top-hat pump, $R$.}
\end{itemize}
Note that the numer of independent concentric rings that can form inside the top-hat depends on its diameter and the Turing pattern wavelength.
Our analysis confirms earlier results on rotating domain walls in optical parametric oscillators and
self-trapped necklace-ring beams in a self-focusing nonlinear Schr\''odinger equation.

Further control over the angular velocity of the pattern can be achieved using vector pumps with orthogonally polarization eigenmodes with good spatial overlap. 
\begin{itemize}
\item{Using cylindrical vector beams, that have eigenmodes with equal and opposite OAM $m$, controlling the relative weightings of the eigenmodes, the ‘bias’, allows the angular velocity to range from $-\frac{2 m}{R^2} \le \omega \le \frac{2 m}{R^2}$}.
\item{Using Poincar\'e beams, that have eigenmodes with different magnitudes  of OAM $m_L, m_R$, the resultant angular velocity is $\omega = (m_L + m_R)/R^2$}.
\end{itemize}

If there is no, or very little, overlap between the modes then concentric Turing pattern rings, each with angular velocity $\omega = 2m_{L,R}/R^2$ will result. This can lead to concentric, counter-rotating Turing patterns creating, for example, an 'optical peppermill'-type structure with full and individual control over the speeds of each counter-rotating ring of pattern. This has potential applications in particle manipulation by using the rotating peak intensities to dipole trap atoms, molecules and small particles. The differential rotation of concentric rings can also be applied to stretching and breaking of cells in a way analogous to optical stretchers \cite{Guck01}.
Finally, rotating Turing patterns can be used to induce circular transport of cold atoms and BEC wavepackets using opto-mechanic nonlinearities instead of Kerr \cite{Baio19}.

\section*{Funding}
We acknowledge support from the Leverhulme Trust Research Project Grant No. RPG-2017-048,
European Training Network ColOpt, which is funded by the European Union (EU) Horizon 2020 programme under the Marie Skłodowska-Curie action, grant agreement 721465, and Engineering and Physical Sciences Research Council DTA Grant No. EP/M506643/1.


\section*{Disclosures}
The authors declare that there are no conflicts of interest related to this article.

\section{Appendix A. Retarded time transformation}
Equation (\ref{LL1DFv2}) is the LL equation on a ring for the field $F(\varphi, t)$. By using the angular velocity (\ref{eqn:omega}) it is possible to introduce retarded time transformations
\begin{equation}
\zeta = \varphi \; ;\;\;\;\;\;\;\;\;\;\;\;\;
\tau = t - \frac{\varphi}{\omega } \, .
\end{equation}
s.t we can write
\begin{eqnarray}
& & \frac{\partial}{\partial \varphi} = \frac{\partial}{\partial \zeta}  - \frac{1}{\omega} \frac{\partial}{\partial \tau} 
\; ;\;\;\;\;\;\;\;\;\;\;\;\;
\frac{\partial}{\partial t} = \frac{\partial}{\partial \tau} \, .
\end{eqnarray}
We can then write the l.h.s of Eq. (\ref{LL1DFv2}) in the retarded time variables:
\begin{eqnarray}
\frac {\partial F}{\partial t} + \omega \frac{\partial F}{\partial \varphi} = 
\frac{\partial F}{\partial \tau}  + \omega \left ( \frac{\partial F}{\partial \zeta} - \frac{1}{\omega} \frac{\partial F}{\partial \tau} \right)
=\omega  \frac{\partial F}{\partial \zeta} \;  \nonumber
\end{eqnarray}
and so we can write Eq. (\ref{LL1DFv2}) as
\begin{eqnarray}
\label{LL1DFv3}
\omega \frac{\partial F}{\partial \zeta} =
P_m-\left[ 1 + i \left( \theta + \frac{m^2}{R^2} \right) \right] F 
+ i \beta |F|^2 F
+ \frac{i} {R^2} \left(\frac{\partial^2 }{\partial \zeta^2} + \frac{1}{\omega^2} \frac{\partial^2 }{\partial \tau^2} 
- 2 \frac{1}{\omega} \frac{\partial }{\partial \zeta} \frac{\partial }{\partial \tau} \right) F . 
\end{eqnarray}

Steady states in the retarded variable $\zeta$, obtained by imposing ${\partial F}/{\partial \zeta} = 0$ in Eq. (\ref{LL1DFv3}), correspond exactly to the rotating solutions (\ref{RotSolut}). Eq. (\ref{LL1DFv3}) can then be used to study the instabilities of the rotating solutions via, for example, appropriate linear stability analyses.


\section{Appendix B. Rotating Turing patterns away from threshold}

Here we investigate azimuthal Turing patterns on a ring due to pump fields carrying OAM well above threshold. From the analysis close to threshold, Turing patterns are spatially modulated structures with wavelength  $\Lambda_c=2\pi/k_c$ where $k_c$ is the critical wavevector given by Eq. (\ref{eqn:kcv2}). We consider spatially modulated solutions of the ring LLE (\ref{LL1D}) of the form
\begin{equation}
\label{rts}
E(\varphi,t) = F(\varphi,t)  e^{i m \varphi} = A[Q(\varphi,t)] \exp{\left\{ i \Phi [Q(\varphi,t)] +  i \psi \right\} } e^{i m \varphi}
\end{equation}
where $\psi$ is a constant phase and $A$ and $\Phi$ are amplitude and phase functions that are periodic in the variable $q = \varphi - \omega t$ and spatially normalised for Turing patterns of wavevectors $k_c$, given by: 
\begin{equation}
Q(\varphi,t) = k_c \, R \, q = k_c \, R \left( \varphi - \omega t \right)
\end{equation}
where $\omega$ is the angular frequency. By replacing (\ref{rts}) in the ring LLE (\ref{LL1DF}) one obtains:
\begin{eqnarray}
\label{partdiff}
&-& k_c R \omega \left( \frac {\partial A} {\partial Q} + i A \frac {\partial \Phi} {\partial Q} \right) 
\exp {\left( i \Phi +  i \psi \right) } =
+ P_m +  \left \{ - \left[ 1 + i \left( \theta + \frac{m^2}{R^2} \right) \right] A + i \beta A^3 \right. \\
&+& k_c^2 \left[  i  \frac {\partial^2 A} {\partial Q^2} - A \frac{\partial^2 \Phi} {\partial Q^2} - 2 \frac{\partial A} {\partial Q} 
\frac{\partial \Phi} {\partial Q} -i A \left( \frac {\partial \Phi} {\partial Q} \right)^2 \right ] 
-  \left . \frac{2 m k_c R}{R^2}  \left( \frac {\partial A} {\partial Q} + i A \frac {\partial \Phi} {\partial Q} \right)   \right  \} \exp {\left( i \Phi +  i \psi \right) } \; .\nonumber
\end{eqnarray}
This demonstrates that above threshold, Turing patterns with an amplitude and a phase that are spatially modulated at the critical wavevector $k_c$ are solutions of  Eq. (\ref{LL1DF}) provided that they rotate at an angular velocity $\omega = 2m/R^2$ and that they satisfy 
\begin{eqnarray}
\label{test}
&& P_m \exp {\left( - i \psi \right) } = \left \{ \left[ 1 + i \left( \theta + \frac{m^2}{R^2} \right) \right] A - i \beta A^3 \right. \\
&&  \left. - k_c^2 \left[  i  \frac {\partial^2 A} {\partial Q^2} - A \frac{\partial^2 \Phi} {\partial Q^2} - 2 \frac{\partial A} {\partial Q} 
\frac{\partial \Phi} {\partial Q} - i A \left( \frac {\partial \Phi} {\partial Q} \right)^2 \right ] \right \} 
\exp {\left( i \Phi  \right) } \nonumber
\end{eqnarray}
We have verified condition (\ref{test}) by integrating Eq. (\ref{LL1D}) well above the threshold of pattern formation and for a variety of OAM indices $m$. In all of these tests, the numerically found rotating Turing patterns are of the form (\ref{rts}) and satisfy condition (\ref{test}) with an error smaller that 2\% up to stationary intensities almost twice the pattern formation threshold.


\end{document}